\documentclass[reprint, amsmath, amssymb, aps, prapplied, citeautoscript,superscriptaddress]{revtex4-2}

\usepackage{graphicx}
\usepackage{dcolumn}
\usepackage{bm}
\usepackage{mathtools}
\usepackage{array}
\usepackage{xcolor}
\usepackage{gensymb}

\begin{document}

\title{Unconventional Field-Like Spin Torques in CrPt$_3$}

\author{Robin Klause}
\altaffiliation[]{Robin Klause and Yuxuan Xiao contributed equally to this work.}
\affiliation{Department of Materials Science and Engineering and Materials Research Laboratory, University of Illinois Urbana-Champaign, Urbana, Illinois 61801, USA}
 
\author{Yuxuan Xiao}
\altaffiliation[]{Robin Klause and Yuxuan Xiao contributed equally to this work.}
\affiliation{Center for Memory and Recording Research, University of California, San Diego, La Jolla, California 92093, USA}
 
\author{Jonathan Gibbons}
\altaffiliation[Now at ]{Western Digital Corporation, San Jose, CA 95119, USA}
\affiliation{Department of Materials Science and Engineering and Materials Research Laboratory, University of Illinois Urbana-Champaign, Urbana, Illinois 61801, USA}
\affiliation{Department of Physics, University of California San Diego, La Jolla, California 92093, USA}

\author{Vivek P. Amin}
\affiliation{Department of Physics, Indiana University, Indianapolis, Indiana 46202, USA}

\author{Kirill D. Belashchenko}
\affiliation{Department of Physics and Astronomy and Nebraska Center for Materials and Nanoscience, University of Nebraska-Lincoln, Nebraska 68588, USA}

\author{Dongwook Go}
\affiliation{Institute of Physics, Johannes Gutenberg University Mainz, 55099 Mainz, Germany}
 
\author{Eric E. Fullerton}
\affiliation{Center for Memory and Recording Research, University of California, San Diego, La Jolla, California 92093, USA}

\author{Axel Hoffmann}%
\affiliation{Department of Materials Science and Engineering and Materials Research Laboratory, University of Illinois Urbana-Champaign, Urbana, Illinois 61801, USA}

\begin{abstract}
The topological semimetal CrPt$_3$ has potential for generating unconventional spin torques due to its ferrimagnetic ordering, topological band structure, and high anomalous Hall effect. CrPt$_3$ exhibits ferrimagnetic behavior only in its chemically ordered phase and is paramagnetic in its chemically disordered phase. By controlling the growth and annealing temperatures, epitaxial films of both chemically ordered and disordered phases of CrPt$_3$ are prepared allowing us to investigate the role of magnetic ordering on unconventional torque generation. We use angle dependent spin-torque ferromagnetic resonance and second harmonic Hall measurements to probe the spin torques generated from epitaxial CrPt$_3$ in CrPt$_3$/Cu/Ni$_{81}$Fe$_{19}$ heterostructures. With current applied along specific directions with respect to the crystal order we reveal unconventional spin torques in both ordered and disordered films. When current flows parallel to the $[1\overline{1}1]$ and $[\overline{1}11]$ directions we observe an unconventional field-like torque that is opposite in sign for the two directions. Our calculations reveal that this unconventional torque originates from an indirect nonlocal spin-orbit torque due to spin scattering at the CrPt$_3$/Cu interface, in addition to symmetry breaking at this interface.
\end{abstract}

\maketitle

\section{Introduction}

Spin-orbit torques play a crucial role in the development of new technology that use the efficient control of magnetization states via charge currents~\cite{Shao2021}. The spin Hall effect is an efficient way of converting a charge current to a spin current and generating spin torques on an adjacent ferromagnetic layer. These spin torques are key to novel information technologies, ranging from power efficient and ultrafast nonvolatile memories~\cite{Garello2019} to new computational paradigms based on magnetic oscillators emulating neuromorphic functionalities~\cite{Hoffmann2022}.

In the conventional spin Hall effect induced spin-orbit torque geometry, a charge current along the in-plane $x$-direction generates a spin-current along the out-of-plane $z$-direction with a spin polarization along the in-plane $y$-direction ($\boldsymbol{\sigma_Y}$) perpendicular to both charge and spin current. The torques that are generated from this spin-polarization are in-plane damping-like torques of the form $\boldsymbol{\tau_{ip}} \propto \textbf{m} \times (\textbf{m} \times \boldsymbol{\sigma_Y})$ and out-of-plane field-like torques of the form $\boldsymbol{\tau_{oop}} \propto \textbf{m} \times \boldsymbol{\sigma_Y}$, with $\textbf{m}$ being the magnetization. These torques can only field-free and deterministically switch an in-plane magnetization parallel to the spin-polarization direction, {\em i.e.}, the $y$-direction.  However, in ordinary spin-transfer torque memory devices, it is well known that in order to have low switching currents while maintaining high thermal stability a perpendicular magnetic anisotropy, {\em i.e.}, along the $z$-direction, is preferred, since this avoids demagnetizing fields during the precessional switching~\cite{Sun2000}.

In recent years, interest in generating spin-currents with a spin-polarization that varies from the conventional geometry has emerged, such that other magnetization directions can be switched deterministically. It has been shown that by symmetry breaking through crystal or magnetic structure, unconventional orientations of spin-polarizations are allowed, such that unconventional torque geometries are possible. In WTe$_2$ the reduced crystal symmetry leads to an out-of-plane spin polarization~\cite{MacNeil2016}, whereas in Mn$_3$Ir and Mn$_3$GaN the non-collinear antiferromagnetic order reduces the symmetry to allow unconventional spin torques~\cite{Holanda2020,Liu2019, Nan2020}. In epitaxial CoPt/CuPt heterostructures, the reduced symmetry at the interface can also result in unconventional spin torques~\cite{Liu2021}. These torques enable field-free deterministic switching of memory devices with perpendicular anisotropies~\cite{Liu2021}, and may also enable novel magnetization dynamics, such as multiple dynamic droplets~\cite{Klause2022}.

Another material that shows potential for generating such torques is the topological semimetal CrPt$_3$ due to its magnetic ordering~\cite{Markou2021}. CrPt$_3$ exhibits ferrimagnetic behavior in its chemically ordered phase and paramagnetic behavior in its chemically disordered phase~\cite{Hellwig2001}. In the ferrimagnetic phase, the magnetic ordering may give rise to unconventional torques by generating a spin polarization with a polarization component collinear to the magnetization~\cite{Gibbons2018,Iihama2018,Baek2018,Safranski2019,Bose2018}.

In this work, we use angle dependent spin-torque ferromagnetic-resonance~\cite{Liu2011} and second harmonic Hall~\cite{Hayashi2014,Avci2014} measurements with current applied along varying crystallographic directions and in both ferrimagnetic and paramagnetic phases to investigate the role of magnetic ordering on unconventional torque generation. We show an unconventional field-like spin torque originating from a spin polarization component along the applied current direction in both ferrimagnetic and paramagnetic CrPt$_3$ thin films. The presence of unconventional torques in both phases suggests that magnetic ordering does not contribute to this torque generation. Instead, our calculations reveal that interlayer spin scattering resulting in an indirect nonlocal spin-orbit torque in combination with symmetry breaking at the CrPt$_3$/Cu interface results in unconventional torque generation.

\section{Methods}
\subsection{Sample growth and characterization}

CrPt$_3$ was grown on single-crystal MgO (220) substrates using a dc magnetron sputtering system with a base pressure of 5 $\times$ 10$^{-8}$~Torr. Cr and Pt targets were co-sputtered using a confocal geometry in an Ar environment with sputtering pressure of 2.7 mTorr. The co-sputtering growth rate was 1.5 \AA/s. CrPt$_3$ films were grown at 500$^\circ$C to achieve epitaxial growth and control the surface roughness. To obtain chemical ordering, the samples were post-annealed \textit{in situ} at 850$^\circ$C for 40 minutes. In a second batch of samples, the post-annealing was skipped to obtain chemically disordered films. The samples were naturally cooled \textit{in situ} followed by Cu and permalloy (Py, Ni$_{81}$Fe$_{19}$, Ni and Fe in wt\%) deposition at room temperature to minimize interfacial mixing. Lastly, an Al$_2$O$_3$ capping layer was deposited to prevent oxidation of the layers below. The Cu spacer layer served to magnetically decouple the CrPt$_3$ and Py layers, while the Py layer was used as a sensing layer for spin-torque measurements. The final sample structures were MgO/CrPt$_3$(15)/Cu(3)/Py(8)/Al$_2$O$_3$(2) for the ordered and MgO/CrPt$_3$(15)/Cu(3)/Py(5)/Al$_2$O$_3$(2) for the disordered sample. The numbers in parenthesis indicate film thicknesses in nm. The thicknesses were chosen to achieve approximately equal resistance of the two layers, such that an equal amount of current flowed through the CrPt$_3$ and Py layers during spin-torque measurements. The thicknesses are based on separate resistivity measurements of CrPt$_3$ and Py patterned full films, and normalized resistivity measurements of Cu(3) on patterned Cu/Ge multilayers. The resistivities were 52~$\mu\Omega$cm for ordered and 82.5~$\mu\Omega$cm for disordered CrPt$_3$.

\begin{figure}
\includegraphics[width=8.6 cm]{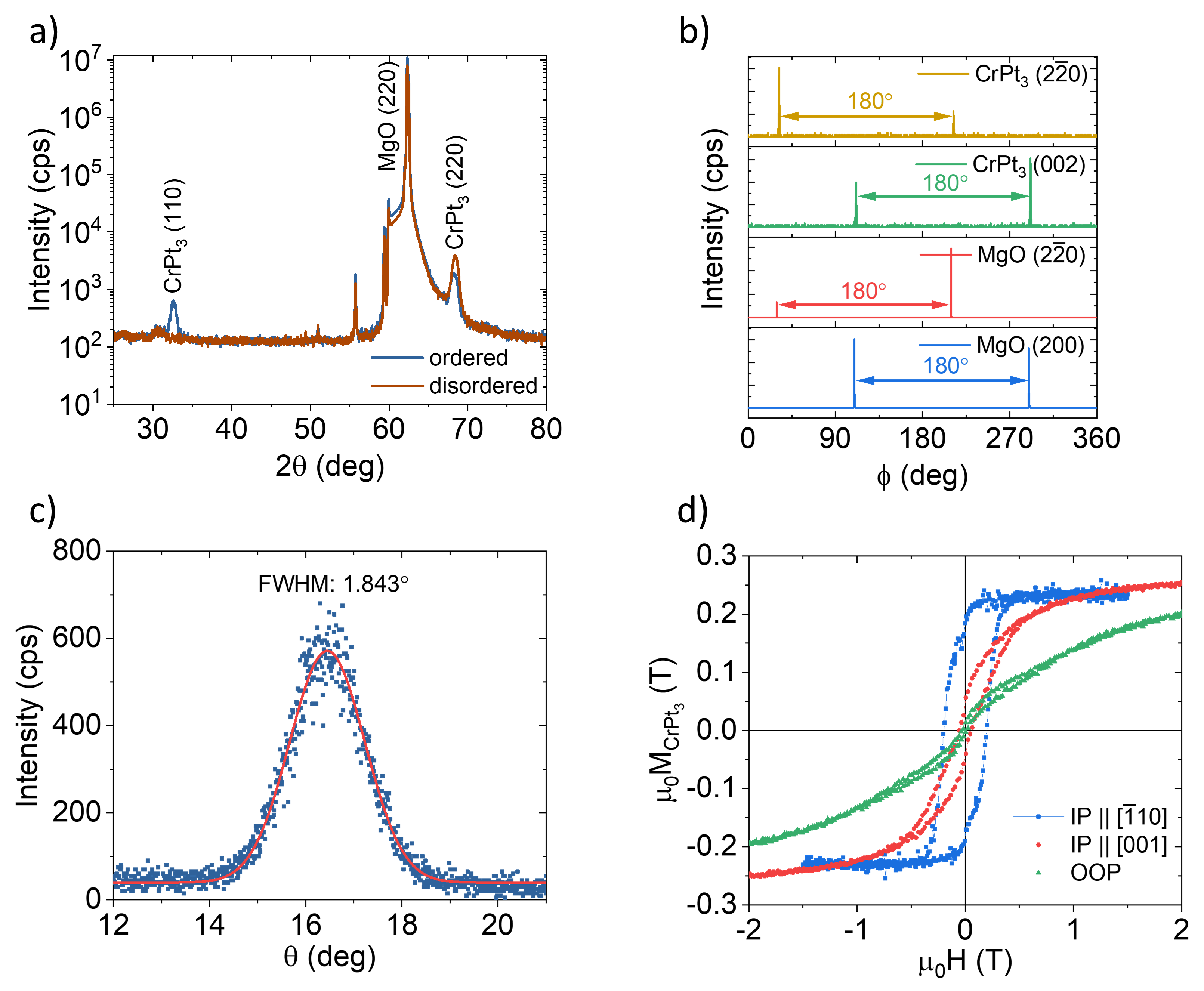}
\caption{(a) Out-of-plane X-ray diffraction measurement of chemically ordered (blue) and disordered (brown) CrPt$_3$, (b) in-plane X-ray diffraction of chemically ordered CrPt$_3$ with 2-fold symmetry aligned with the MgO (220) substrate (c) rocking curve over CrPt$_3$ (110) diffraction peak, and (d) room temperature vibrating sample magnetometry of chemically ordered CrPt$_3$ with magnetic field along the in-plane $[\overline{1}10]$ (blue), $[001]$ (red), and out-of-plane (green) directions.}
\label{Characterization}
\end{figure}

The chemical ordering of CrPt$_3$ was confirmed using out-of-plane X-ray diffraction as shown in Fig.~\ref{Characterization}(a). While both films show a (220) peak, chemically ordered CrPt$_3$ exhibits a characteristic (110) peak whereas disordered CrPt$_3$ does not since the (110) peak is forbidden in disordered fcc crystals. As shown in Fig.~\ref{Characterization}(c), the full width at half maximum of the $\theta$ scan over CrPt$_3$~(110) is 1.843$^\circ$, indicating good crystalline quality. To confirm epitaxy, in-plane $\phi$ scans were performed, which show two-fold symmetry and alignment with the MgO substrate, see [Fig.~\ref{Characterization}(b)]. Note, that the disordered CrPt$_3$ films are chemically and not structurally disordered so they have the same fcc crystal structure as the chemically ordered films. In the chemically ordered films the Cr atoms always occupy the corner sites, while the Pt atoms always occupy the face-center sites of the fcc structure. In the chemically disordered films the Cr and Pt atoms can occupy any site as long as the Cr:Pt 1:3 stoichiometry is met. This is shown in Fig.~\ref{Sample_setup}(a).

Magnetization loops of ordered CrPt$_3$ were measured using vibrating sample magnetometry with magnetic fields applied along different crystal directions as shown in Fig.~\ref{Characterization}(d). The measurements show that the magnetic easy axis is along $[\overline{1}10]$ with a coercive field of $\mu_0 H_{c}=0.3$~T. The (110) growth direction was specifically chosen, such that the magnetic easy axis of CrPt$_3$ lies in plane and current can be applied along varying in-plane directions with respect to the easy axis to understand the role of the magnetic order on the spin-torque geometry. The large coercive field allows for the magnetization of CrPt$_3$ to be set using a correspondingly large magnetic field and not be changed during spin-torque ferromagnetic resonance measurements with a smaller magnetic field, where only the Py magnetization is rotated. The maximum magnetic field used during measurements was $\mu_0 H_{ext}=0.135$~T.

\subsection{Spin-torque measurements}

\begin{figure}
\includegraphics[width=8.6 cm]{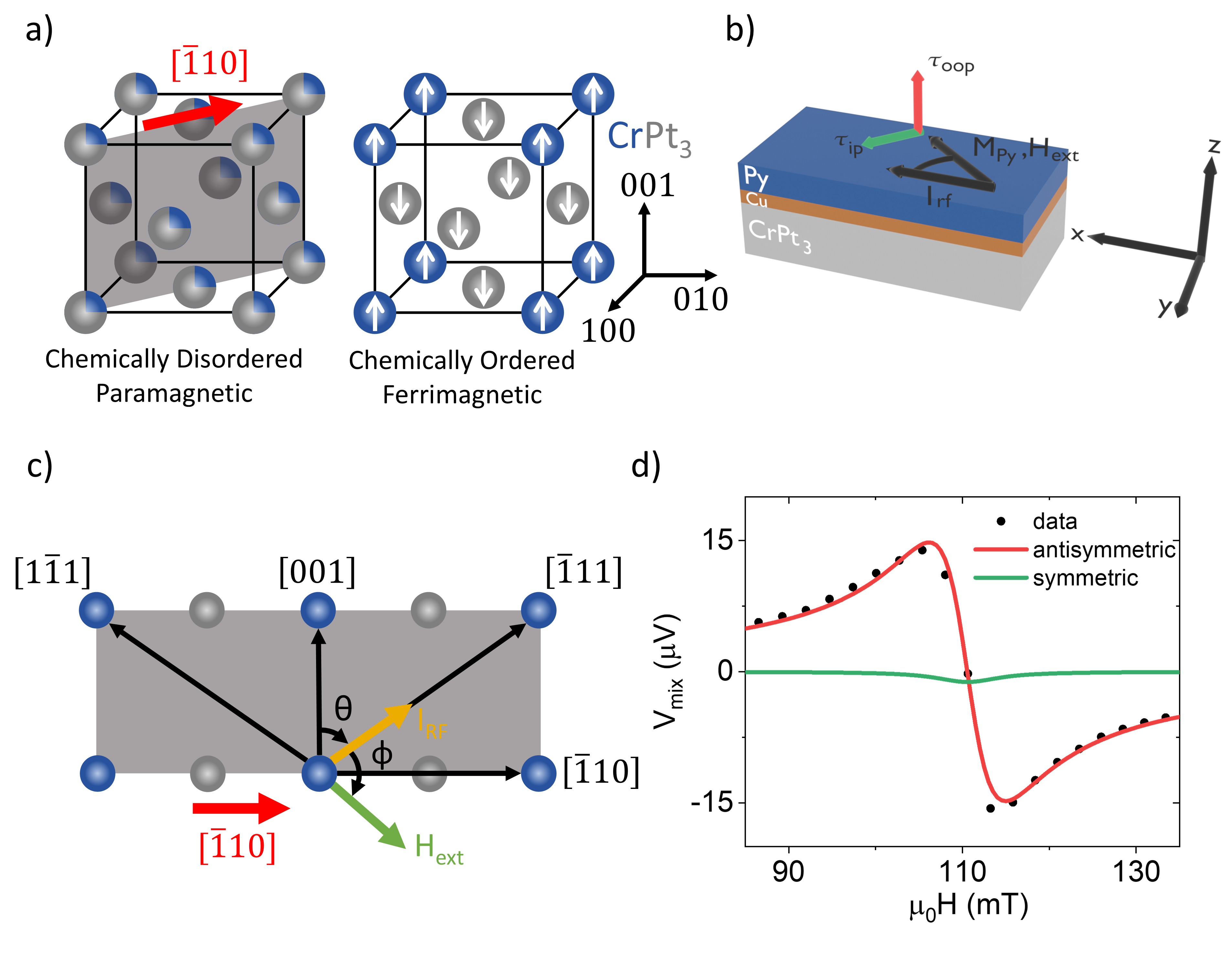}
\caption{(a) Crystal and magnetic structure of chemically ordered (right) and disordered (left) CrPt$_3$, where blue spheres represent Cr and gray spheres represent Pt atoms. The gray plane corresponds to the film plane. (b) Schematic of the device structure used for spin-torque ferromagnetic resonance measurements with $\tau_{oop}$ corresponding to out-of-plane torques manifested in the antisymmetric and $\tau_{ip}$ corresponding to in-plane torques manifested in the symmetric Lorentzian component. $\phi$ is the angle between the {\em rf} current ($I_{rf}$) and the magnetization ($M_{Py}$) that is determined by the magnetic field ($H_{ext}$). (c) Schematic of the four crystal orientations in which rf current is applied. $\theta$ represents the angle between the $[001]$ direction and the applied current ($I_{rf}$, yellow arrow) and $\phi$ the angle between the applied current and the magnetic field ($H_{ext}$, green arrow). The red arrow shows the magnetic easy axis direction that is only present in the ordered sample. (d) Exemplary field sweep at a single angle. The resonance lineshape can be decomposed into an antisymmetric (red) and symmetric (green) component.}
\label{Sample_setup}
\end{figure}

To perform spin-torque measurements, the multilayer films were patterned using photolithography and subsequent ion milling into ($10 \times 60$)~$\mu$m wires for spin-torque ferromagnetic resonance and ($10 \times 70$)~$\mu$m Hall bars for second harmonic Hall measurements. In a second lithography step and subsequent sputtering of Ti and Au, coplanar waveguides and contact pads were patterned to electrically contact the devices. Devices were patterned with wires and Hall bars along different crystal directions and thereby different directions with respect to the magnetic easy axis. By comparing spin torques generated from different device orientations the role of the magnetic order and crystal structure can be investigated. Devices were patterned along the $[001]$ ($\theta=0^{\circ}$), $[\overline{1}10]$ ($\theta=90^{\circ}$), $[\overline{1}11]$ ($\theta=+54.7^{\circ}$), and $[1\overline{1}1]$ ($\theta=-54.7^{\circ}$) directions. $\theta$ represents the angle between the $[001]$ crystal direction and the current direction as shown in Fig.~\ref{Sample_setup}(c). Prior to the spin-torque measurements of the ordered CrPt$_3$ devices, the sample was placed in a $0.5\;$T magnetic field along the $[\overline{1}10]$ direction to saturate the magnetization of the CrPt$_3$ along its magnetic easy axis. Due to the large coercivity, the magnetic fields used during the measurements were not strong enough to switch the magnetization direction of the CrPt$_3$. Together with the different device orientations this allowed us to study the effect of the CrPt$_3$ magnetization on the spin-torque geometry.

Spin-torque ferromagnetic-resonance measurements were used to characterize the spin torques generated in CrPt$_3$. In these measurements a 7$\;$GHz 10$\;$dBm {\em rf} current is applied through the device which generates an {\em rf} spin current in the CrPt$_3$ that is injected through the Cu into the Py. There, the spins exert a torque on the magnetization of the Py, and together with Oersted fields drive the magnetization into precession around an in-plane applied magnetic field when the magnetic field matches the resonance field of the Py. This leads to an {\em rf} resistance modulation due to the anisotropic magnetoresistance in Py. The mixing of the {\em rf} resistance and the applied {\em rf} current generates a {\em dc} mixing voltage, which is measured using a lock-in amplifier.

The magnetic field is swept through the resonance condition of the Py to measure the resonance line shape, which is composed of a symmetric and antisymmetric Lorentzian function and can be written as

\begin{equation}
\begin{aligned}
    V_{mix} & = S \frac{\Delta H^2}{(H_{ext}-H_0)^2 + \Delta H^2}  \\
            & + A \frac{\Delta H (H_{ext}-H_{R})}{(H_{ext}-H_{R})^2 + \Delta H^2}
    \end{aligned}
\label{Lorentzian}
\end{equation}

Here, $\Delta H$ is the half width at half maximum of the resonance linewidth, $H_{ext}$ is the applied magnetic field, $H_{R}$ is the Py resonance field, and \textit{S} and \textit{A} are the symmetric and antisymmetric Lorentzian components, respectively, that measure the size of the spin torques. An exemplary fit to a field sweep is shown in Fig.~\ref{Sample_setup}(d). The antisymmetric component corresponds to out-of-plane torques, while the symmetric component corresponds to in-plane torques [$\tau_{oop}$ and $\tau_{ip}$ respectively in Fig.~\ref{Sample_setup}(b)].

To fully analyze the spin-torque symmetry, the magnetic field is rotated through a 360$^{\circ}$ in-plane angle to vary the angle between the applied current and Py magnetization [denoted by $\phi$ as shown in Fig.~\ref{Sample_setup}(c)], and a magnetic field sweep is performed every 5$^{\circ}$. The size of the symmetric and antisymmetric components depends on the relative orientation between the spin-polarization and Py magnetization, such that the spin torque and thereby spin-polarization direction can be extracted from the angular dependence of the voltage signals.

Additionally, second harmonic Hall measurements were done to confirm the spin-torque geometry. For these measurements an {\em ac} current with an amplitude of $5\;$mA and a frequency of $1311\;$Hz was passed through a Hall bar device and the transverse voltage was measured at the second harmonic frequency. The {\em ac} current causes an {\em ac} tilt in the magnetization of the Py, which causes an oscillating change in the planar Hall resistance at twice the frequency of the applied current. The change in resistance mixes with the applied current to generate a voltage, that is measured using a lock-in amplifier. In addition, a $0.135$~T magnetic field is used to rotate the magnetization in-plane and a voltage is measured every 5$^{\circ}$. Similar to the spin-torque ferromagnetic resonance measurements, the size of the voltage depends on the relative orientation of the spin polarization and magnetization, so by looking at the magnetic field direction dependence of the voltage, the spin polarization can be extracted.

\section{Results}

\begin{figure}
\includegraphics[width=8.6 cm]{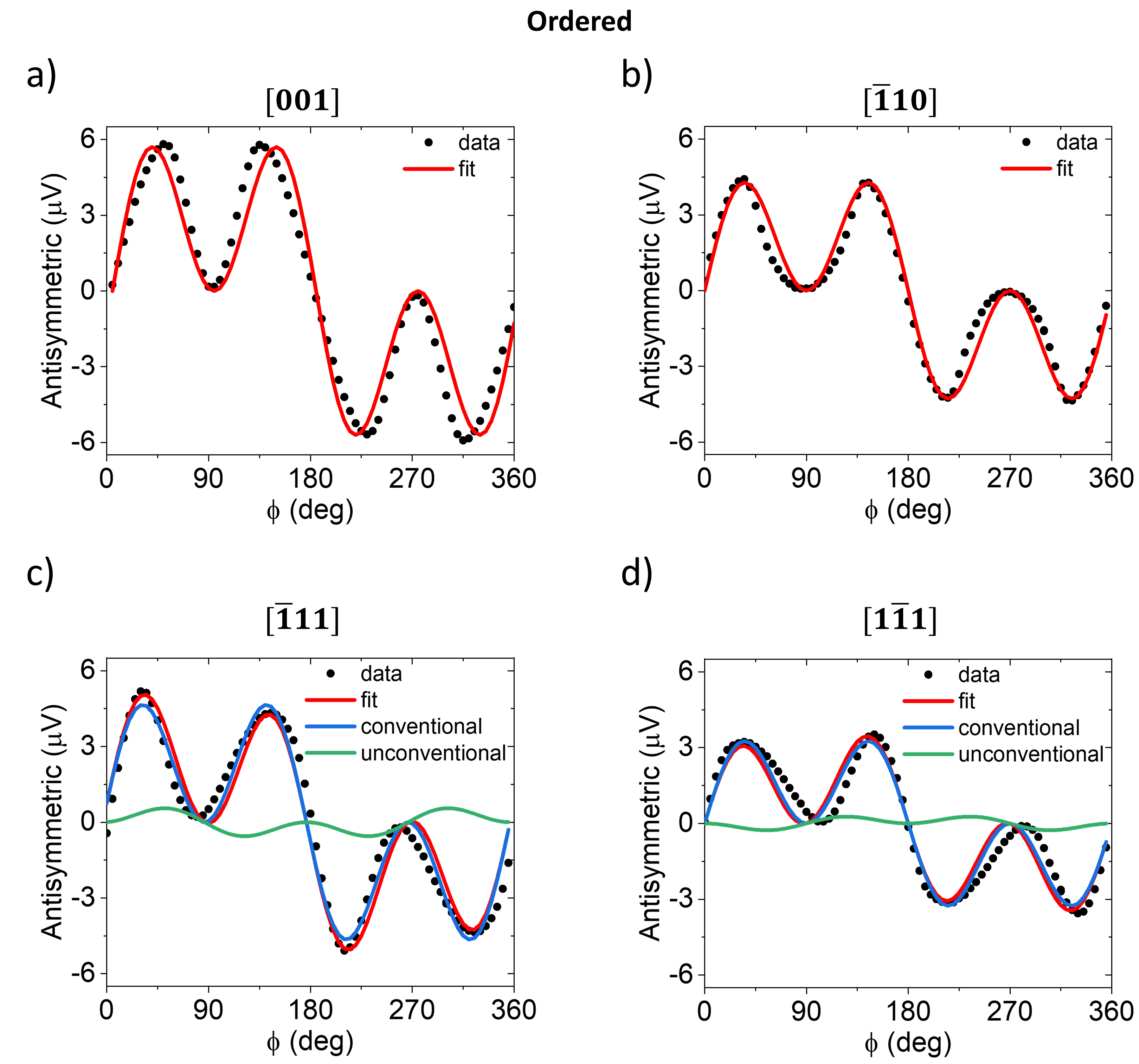}
\caption{Antisymmetric Lorentzian component of the resonance signal as a function of the angle between the current and magnetic field for the ordered CrPt$_3$ devices with current along the (a) $[001]$, (b) $[\overline{1}10]$, (c) $[\overline{1}11]$, and (d) $[1\overline{1}1]$ crystal direction of CrPt$_3$. The red line shows the fit of the data in black circles. For (c) and (d) the blue (green) lines show the contribution from conventional (unconventional) torque components.}
\label{STFMR_Ordered}
\end{figure}

\begin{figure}
\includegraphics[width=8.6 cm]{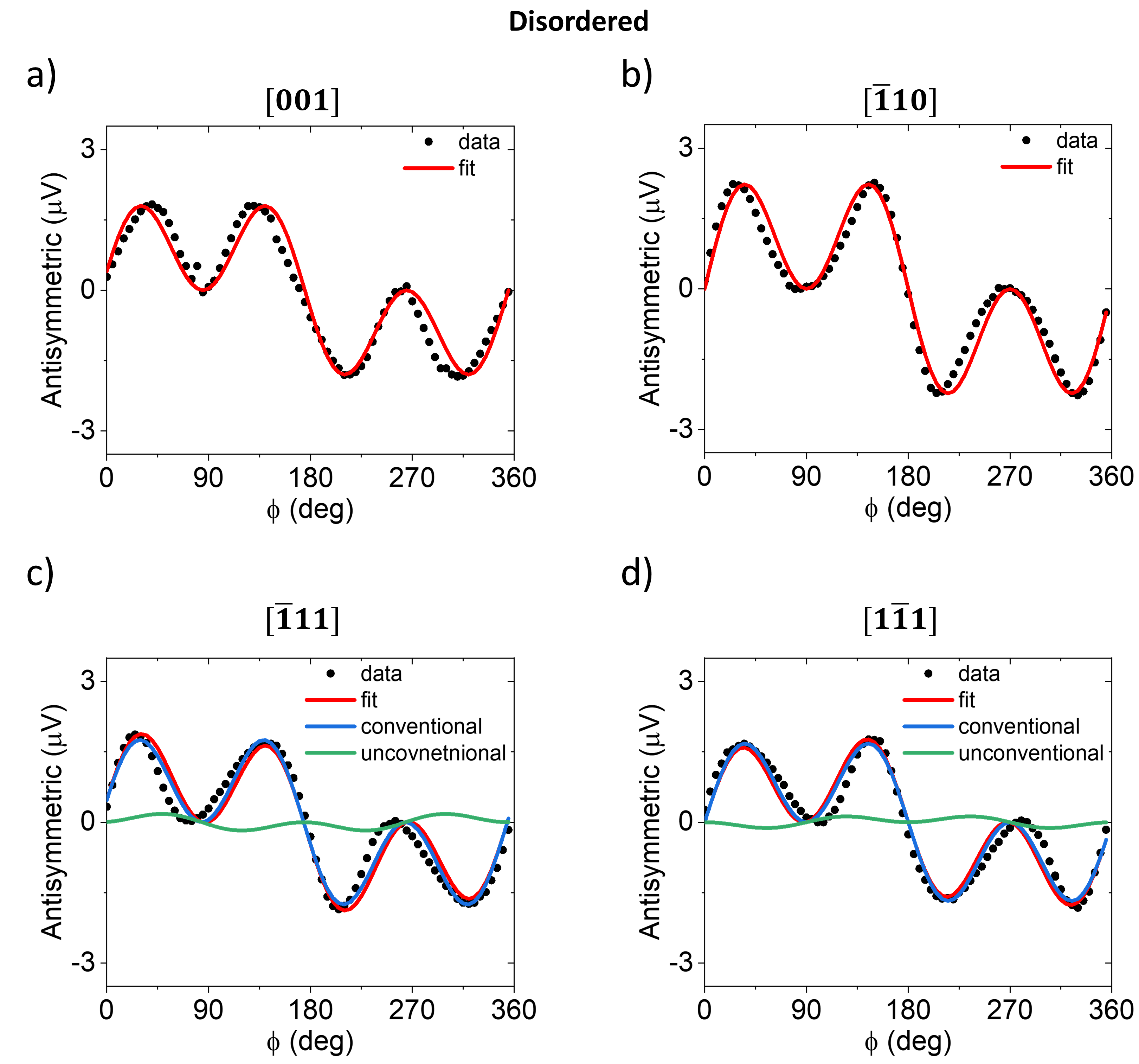}
\caption{Antisymmetric Lorentzian component of the resonance signal as a function of the angle between the current and magnetic field for the disordered CrPt$_3$ devices with current along the (a) $[001]$, (b) $[\overline{1}10]$, (c) $[\overline{1}11]$, and (d) $[1\overline{1}1]$ crystal direction of CrPt$_3$. The red line shows the fit of the data in black circles. For (c) and (d) the blue (green) lines show the contribution from conventional (unconventional) torque components.}
\label{STFMR_Disordered}
\end{figure}

\begin{table}
\caption{Fitting parameters of the spin-torque ferromagnetic resonance ($A_{FL}^Y$ and $A_{FL}^X$) and second harmonic Hall ($V_{FL}^Y$, $V_{DL}^Y$, and $V_{FL}^X$) data in $\mu$V.}
\label{table}
\begin{ruledtabular}
\begin{tabular}{c|cccc}
    Ordered & $[001]$ & $[\overline{1}10]$ & $[\overline{1}11]$ & $[1\overline{1}1]$ \\
    \hline
    $A_{FL}^Y$ & 7.41 & 5.56 & 6.04 & 4.23 \\  
    $A_{FL}^X$ & 0 & 0 & 0.72 & -0.35 \\
    $V_{FL}^Y$ & 0.134 & 0.094 & 0.109 & 0.138 \\
    $V_{DL}^Y$ & 0.015 & 0.010 & 0.018 & 0.017 \\
    $V_{FL}^X$ & 0 & 0 & 0.034 & -0.036 \\
    \hline\hline
    Disordered & $[001]$ & $[\overline{1}10]$ & $[\overline{1}11]$ & $[1\overline{1}1]$ \\
    \hline
    $A_{FL}^Y$ & 2.35 & 2.90 & 2.28 & 2.18 \\
    $A_{FL}^X$ & 0 & 0 & 0.23 & -0.16 \\
    $V_{FL}^Y$ & 0.142 & 0.148 & 0.174 & 0.164 \\
    $V_{DL}^Y$ & 0.009 & 0.011 & 0.011 & 0.013 \\
    $V_{FL}^X$ & 0 & 0 & 0.044 & -0.030 \\
\end{tabular}
\end{ruledtabular}
\end{table}

Figure~\ref{STFMR_Ordered} shows the angle dependent antisymmetric mixing voltage component of the ordered ferrimagnetic CrPt$_3$ sample. The signal from the devices oriented along $[001]$ (0$^{\circ}$) and $[\overline{1}10]$ (90$^{\circ}$)  [Figs.~\ref{STFMR_Ordered}(a) and (b), respectively] can be fit with $A=A_{FL}^Y\mathrm{sin}(2\phi)\mathrm{cos}(\phi)$, where $A_{FL}^Y$ is related to the strength of the field-like torque originating from conventional $y$-direction polarized spins and Oersted fields. To fit the signal from the devices oriented along $[\overline{1}11]$ (+54.7$^{\circ}$) and $[1\overline{1}1]$ (-54.7$^{\circ}$) an additional $A_{FL}^X\mathrm{sin}(2\phi)\mathrm{sin}(\phi)$ component needs to be added, where $A_{FL}^X$ is related to the strength of a field-like torque originating from unconventional spins polarized parallel to the current direction. The sign of this component is opposite in the $[\overline{1}11]$ and $[1\overline{1}1]$ devices.

Figure~\ref{STFMR_Disordered} shows the angle dependent antisymmetric mixing voltage for the chemically disordered paramagnetic CrPt$_3$ sample. Here, the $[001]$ and $[\overline{1}10]$ devices [Figs.~\ref{STFMR_Disordered}(a) and (b), respectively] can again be fit with the conventional $A=A_{FL}^Y\mathrm{sin}(2\phi)\mathrm{cos}(\phi)$ term, while the $[\overline{1}11]$ and $[1\overline{1}1]$ devices require an additional $A_{FL}^X\mathrm{sin}(2\phi)\mathrm{sin}(\phi)$ term just like the devices in the ordered sample. Table~\ref{table} shows a summary of the fitting parameters used for each sample and device. We performed the same spin-torque ferromagnetic-resonance measurements on a Pt/Py control sample and see no unconventional torque component, as shown in the Supplemental Material~\cite{Supplemental} ruling out measurement artifacts.

\begin{figure}
\includegraphics[width=8.6 cm]{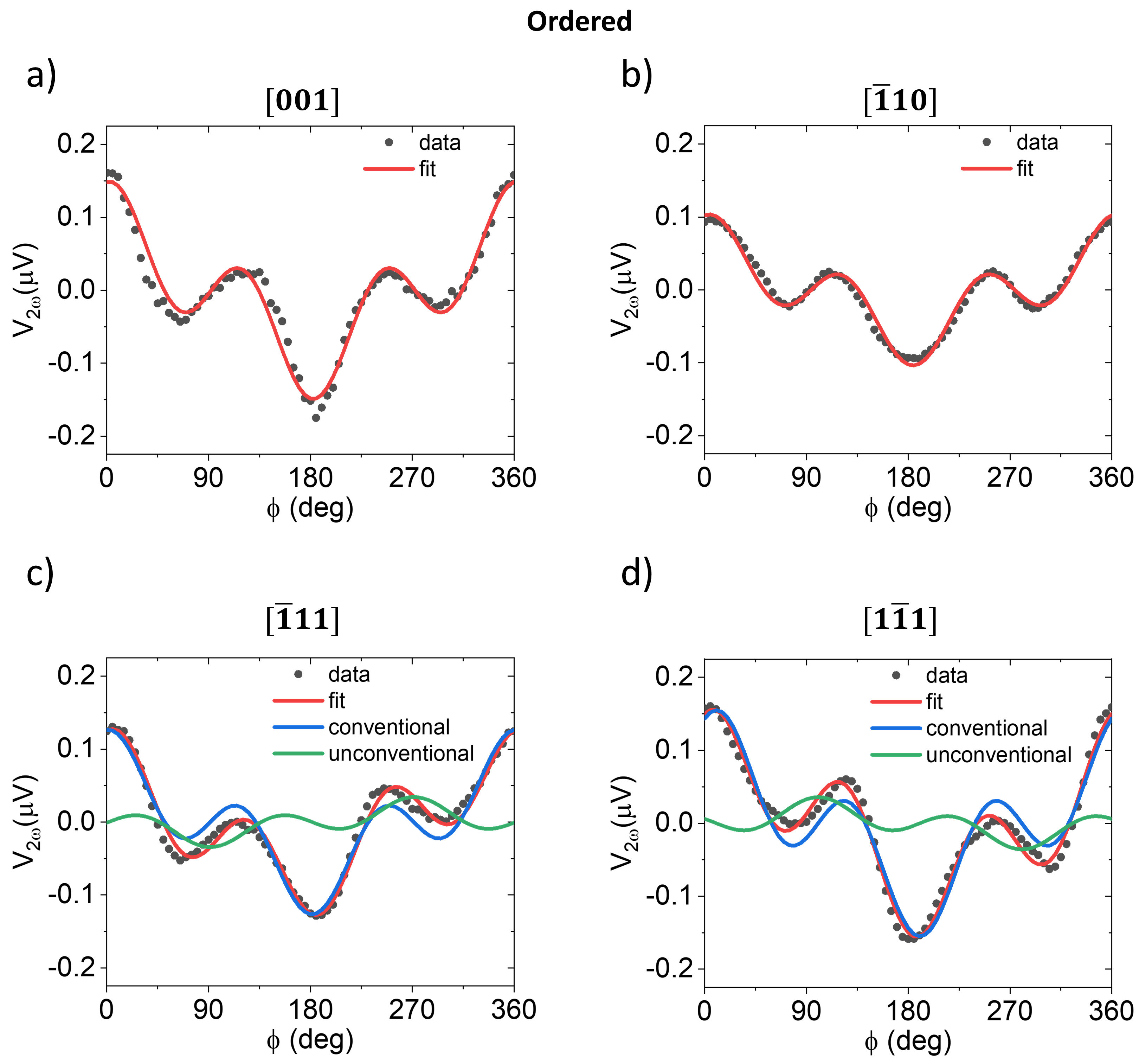}
\caption{Second harmonic Hall voltage as a function of the angle between the current and magnetic field for the ordered CrPt$_3$ Hall bar devices with current along the (a) $[001]$, (b) $[\overline{1}10]$, (c) $[\overline{1}11]$, and (d) $[1\overline{1}1]$ crystal direction of CrPt$_3$. The red line shows the fit of the data in black circles. For (c) and (d) the blue (green) lines show the contribution from conventional (unconventional) torque components.}
\label{SHH_Ordered}
\end{figure}

\begin{figure}
\includegraphics[width=8.6 cm]{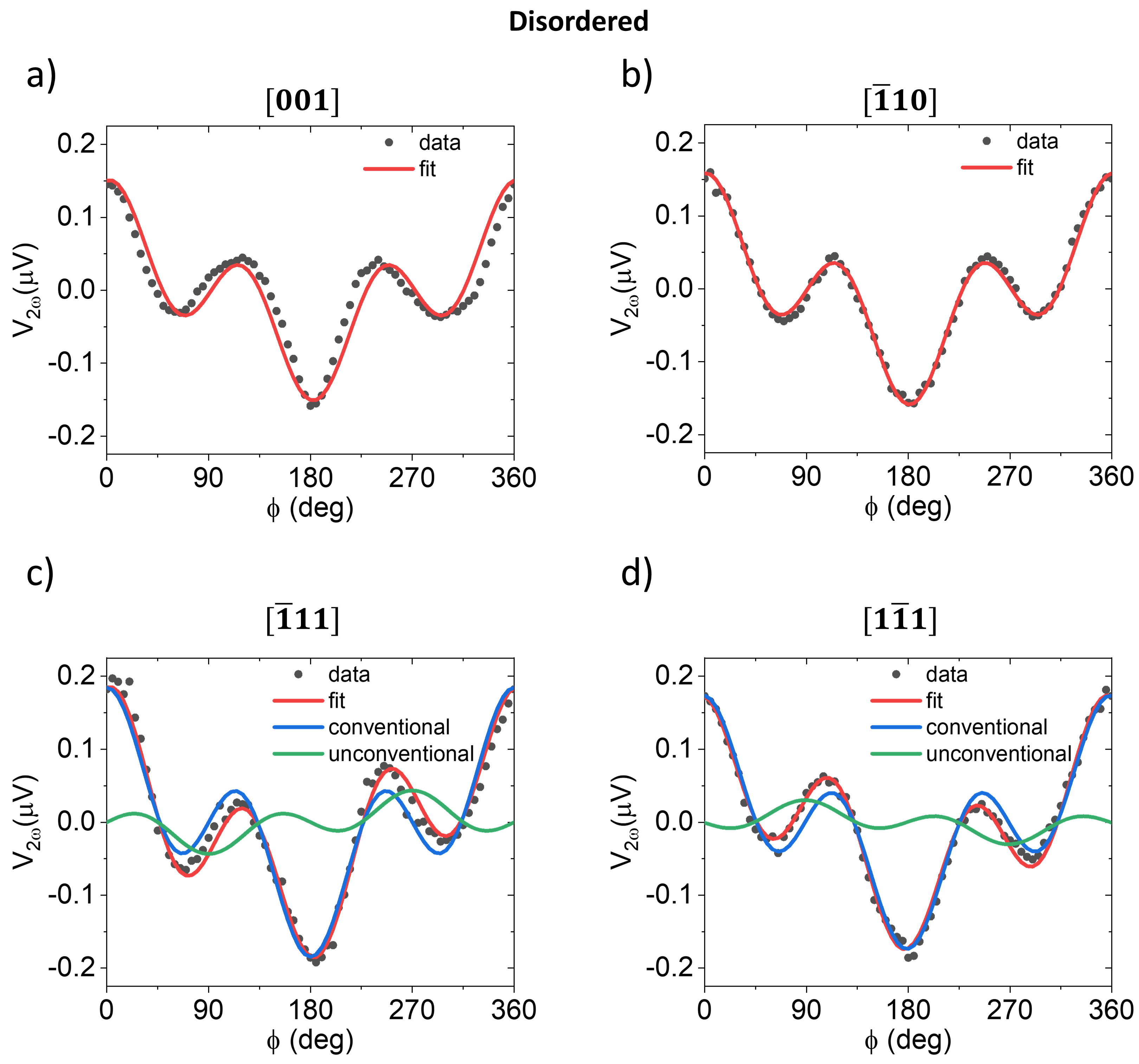}
\caption{Second harmonic Hall voltage as a function of the angle between the current and magnetic field for the disordered CrPt$_3$ Hall bar devices with current along the (a) $[001]$, (b) $[\overline{1}10]$, (c) $[\overline{1}11]$, and (d) $[1\overline{1}1]$ crystal direction of CrPt$_3$. of CrPt$_3$. The red line shows the fit of the data in black circles. For (c) and (d) the blue (green) lines show the contribution from conventional (unconventional) torque components.}
\label{SHH_Disordered}
\end{figure}

The results of the second harmonic Hall measurements are shown in Figures~\ref{SHH_Ordered} and~\ref{SHH_Disordered} for the ordered and disordered samples, respectively. Similar to the spin-torque ferromagnetic resonance measurements, there is no qualitative difference in the geometry of the spin torques in the ordered and disordered films. Devices with current along $[001]$ (0$^{\circ}$) and $[\overline{1}10]$ (90$^{\circ}$) can be fit with conventional field-like and damping-like torque terms that take the form $V_{2\omega}=V_{FL}^Y\mathrm{cos}(\phi)\mathrm{cos}(2\phi)+V_{DL}^Y\mathrm{cos}(\phi)$. Here $V_{2\omega}$ is the second harmonic Hall voltage, and $V_{FL}^Y$ and $V_{DL}^Y$ are related to the strength of the conventional field-like and damping-like torques due to $y$-direction polarized spins and Oersted fields.

Devices with current along $[\overline{1}11]$ ($+54.7^{\circ}$) and $[1\overline{1}1]$ ($-54.7^{\circ}$) need an additional $V_{2\omega}=V_{FL}^X\mathrm{sin}(\phi)\mathrm{cos}(2\phi)$ term. $V_{FL}^X$ is related to the strength of the unconventional field-like torque due to $x$-direction polarized spins. The sign of this term is opposite for the $[\overline{1}11]$ and $[1\overline{1}1]$ devices as shown in Table~\ref{table} along with the other fitting parameters. Note that for these measurements $\phi$ includes an angle offset term that accounts for the slight misalignment between the current and field directions, which is typically no larger than $\pm$10$^{\circ}$. Additionally, each device has a voltage offset that was subtracted for better depiction and comparison of the data.

The spin-torque ferromagnetic resonance and second harmonic Hall measurements are in qualitative agreement. Both show an unconventional field-like component in devices with current along the $[\overline{1}11]$ and $[1\overline{1}1]$ crystal directions that are opposite in sign for the two directions. Devices with current along the $[001]$ and $[\overline{1}10]$ show no unconventional component.

If the unconventional spin torque originated from the magnetic ordering of the CrPt$_3$ the spin polarization is expected to have a component along the magnetization direction. This would result in a purely conventional spin polarization in the device with current along the $[001]$ direction since the magnetization of the CrPt$_3$ is perpendicular to the current direction. With current along the $[\overline{1}10]$ direction, the spin polarization should have the largest unconventional spin-polarization component along the current direction and collinear to the magnetization. This is not the case for our measurements where we only observe an unconventional spin polarization component with current along the $[\overline{1}11]$ and $[1\overline{1}1]$ directions. This alone suggests that the origin of the unconventional torque is not the magnetic ordering of the CrPt$_3$. Additionally, since we observe an unconventional torque component even in disordered paramagnetic CrPt$_3$, where the magnetic ordering is absent, the unconventional torque cannot come from the magnetic ordering.


Unconventional spin-orbit torques arise when the system breaks the appropriate mirror plane symmetry, where the mirror plane is parallel to the electric field $\bold{E}$ and the out-of-plane vector $\bold{z}$. Breaking this symmetry enables spin currents with flow along $\bold{z}$ and spin polarization along $\bold{E}$ and $\bold{z}$. In the following, we denote spin currents by $Q_{ij}$ for {$i,j\in[X,Y,Z]$}, where $i$ is the spin flow direction and $j$ is the spin polarization direction. Assuming that the electric field points along $X$, the unconventional spin-orbit torques may arise from the spin currents $Q_{ZX}$ and $Q_{ZZ}$. 

\begin{figure*}
\includegraphics[width=1\textwidth]{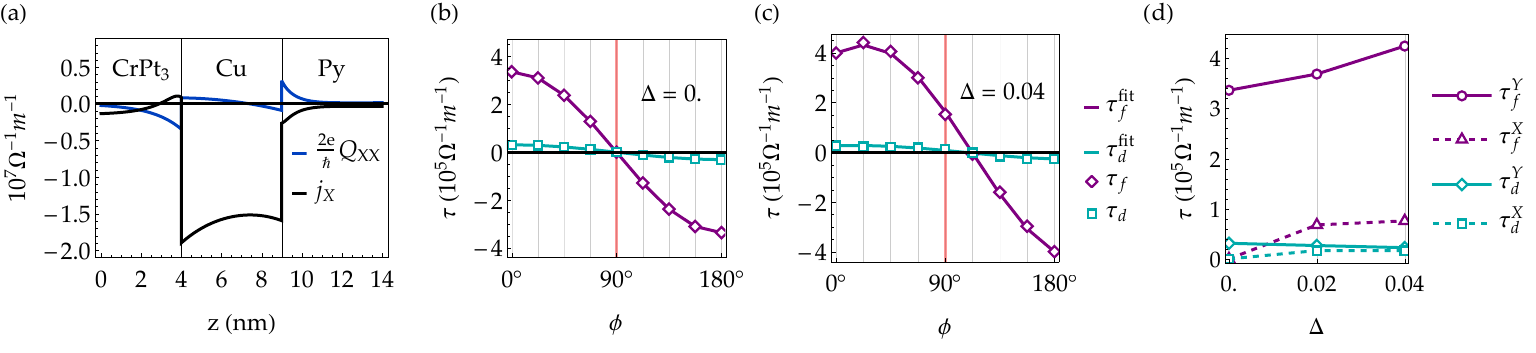}
\caption{Spin-dependent Boltzmann calculations of the CrPt$_3$/Cu/Py trilayer with quantum coherent boundary conditions. (a) In-plane charge current $j_X$ and spin-polarized current $Q_{XX}$ versus out-of-plane position $z$, where $\hat{\bold{M}}_\text{Py} = \hat{\bold{x}}$. (b)-(c) damping-like (cyan curves) and field-like (purple curves) torques versus in-plane magnetization direction $\phi$ for different $\Delta$ values, where the fitted curves are defined using Eq. \ref{FitEq}. Unconventional spin-orbit torques refer to nonvanishing torques when $\hat{\bold{M}}_\text{Py}~||~\hat{\bold{z}} \times \bold{E}$ (here at $\phi = 90\degree$), and are seen (as expected by symmetry arguments) when $\Delta \neq 0$. (d) Conventional (Y) and unconventional (X) torques, defined by Eqs. (\ref{DefConv})-(\ref{DefUnconv}) as a function of $\Delta$. For CrPt$_3$/Cu interfaces with broken mirror plane symmetry (i.e., $\Delta \neq 0$), both the conventional and unconventional field-like torques are larger than the damping-like torques.}
\label{Fig_Boltzmann}
\end{figure*}

In the samples measured, the CrPt$_3$/Cu interface breaks the appropriate mirror plane symmetry while the Py layer, which receives the torque, does not. Thus, the microscopic mechanism for unconventional torques likely involve interlayer scattering. If the CrPt$_3$/Cu layer or the CrPt$_3$/Cu interface generates the spin current $Q_{ZX}$, a damping-like torque $\boldsymbol{\tau} \propto \bold{M}_\text{Py} \times (\bold{x} \times \bold{M}_\text{Py})$ results in Py via the spin transfer mechanism. An unconventional field-like torque, given by $\boldsymbol{\tau} \propto \bold{M}_\text{Py} \times \bold{x}$, results if the incident spin current's polarization, initially along $\bold{x}$, is rotated about the magnetization, such that the polarization now has a component along $\bold{M}_\text{Py} \times \bold{x}$. The resulting spin transfer torque from the new component is $\boldsymbol{\tau} \propto \bold{M}_\text{Py} \times [(\bold{M}_\text{Py} \times \bold{x}) \times \bold{M}_\text{Py}] = \bold{M}_\text{Py} \times \bold{x}$, which is field-like. 

The rotation of the spin current polarization, which results in field-like torques, can occur in two ways. First, if the Cu/Py interface has a nonvanishing imaginary part of the spin mixing conductance, the incident spin current's polarization is rotated about the magnetization upon reflection. However, for this mechanism to explain a ratio of field-like to damping-like torque greater than one, as seen in the experimental data, the ratio of imaginary to real part of the spin mixing conductance would also be greater than one, which is extremely unlikely at metallic interfaces. The other mechanism, called an \emph{indirect nonlocal spin-orbit torque} \cite{Amin2023}, results from a combination of interlayer scattering and spin-to-spin conversion at the interface. 

In the indirect, nonlocal mechanism, the in-plane, spin-polarized current in Py ``leaks" through the spacer layer into the CrPt$_3$ layer. This leakage is the well-known mechanism behind the current-in-plane giant magnetoresistance and is readily modeled by the spin-dependent Boltzmann equation, as seen in Fig. \ref{Fig_Boltzmann}(a). The leaked, in-plane, spin current is polarized along $\bold{M}_\text{Py}$. At the CrPt$_3$/Cu interface, spin-orbit scattering results in spin-to-spin conversion, where the leaked, in-plane spin current is converted into an out-of-plane spin current with spin polarization $\bold{M}_\text{Py} \times \bold{f}$, where the vector $\bold{f}$ is a function of the tensor elements that describe spin-to-spin conversion at the interface. The spin-to-spin conversion can result from spin-orbit precession~\cite{Amin2018,Amin2020} at the CrPt$_3$/Cu interface or from spin swapping~\cite{Lifshits2009} in the CrPt$_3$ layer. The emitted spin current then flows through Cu into Py and exerts a spin-transfer torque given by $\boldsymbol{\tau} \propto \bold{M}_\text{Py} \times [(\bold{M}_\text{Py} \times \bold{f}) \times \bold{M}_\text{Py}] = \bold{M}_\text{Py} \times \bold{f}$. The vector $\bold{f}$ has a nonvanishing $x$ and $y$ components in general (these components are quantified in free-electron models in~\cite{Amin2023}), thus resulting in both conventional and unconventional field-like torques.

We use a multilayer, spin-dependent Boltzmann formalism with quantum coherent boundary conditions to demonstrate that indirect, nonlocal spin-orbit torques plausibly explain the experimental results. Solving the multilayer Boltzmann equation yields the nonequilibrium charge distribution function $g_c(z, \bold{k})$ and nonequilibrium spin distribution function $g_{s,l}(z, \bold{k})$ for $l \in [x,y,z]$, which are used to compute the spin/charge currents and spin torques. We model the CrPt$_3$/Cu/Py trilayer assuming a free electron dispersion relation within each layer and use the spin-dependent relaxation time approximation to construct the collision integrals. All relevant details of the theoretical formalism are provided in~\cite{Amin2023}, though here we have added diffusive scattering at the interfaces to model disorder.

The interfacial scattering potential is given by:
\begin{align}
\label{IntPot}
V(\bold{r},\bold{k}) &= 
\frac{\hbar^2 k_\text{F}}{2m} \delta(z) \Big{(} 
\mu_0 + \mu_\text{ex} \boldsymbol{\sigma} \cdot \hat{\bold{M}}_\text{Py} \nonumber \\ 
&+ (\mu_\text{R} + \Delta/2) \sigma_x k_y
- (\mu_\text{R} - \Delta/2) \sigma_y k_x \Big{)}
\end{align}
where $k_\text{F}$ is the Fermi momentum, $m$ is the effective electron mass, and $\mu_0$, $\mu_\text{ex}$, and $\mu_\text{R}$ are dimensionless parameters specifying the strength of the spin-independent potential barrier, exchange interaction, and Rashba spin-orbit coupling respectively. The dimensionless parameter $\Delta$ specifies the degree of mirror-plane symmetry-breaking at the interface, where $\Delta = 0$ corresponds to an interface with the mirror plane symmetry intact. Aside from spin-flip scattering in the bulk layers (which causes spin diffusion but not spin current generation), spin-orbit coupling is turned off everywhere in the system except the CrPt$_3$/Cu interface. Thus, we isolate nonlocal spin-orbit torques due to interlayer scattering, removing all contributions from bulk spin current generation (like the spin Hall effect, spin anomalous Hall effect, and the magnetic spin Hall effect) and inverse spin galvanic effects at the Cu/Py interface.

Fig. \ref{Fig_Boltzmann}(a) shows the in-plane charge current $j_X$ and spin-polarized current $Q_{XX}$ as a function of out-of-plane position $z$, where $\hat{\bold{M}}_\text{Py} = \hat{\bold{x}}$. The spin-polarized current ``leaks" into the CrPt$_3$ layer, even changing signs due to spin-dependent scattering at the interfaces. Figs. \ref{Fig_Boltzmann}(b)-(c) show the damping-like torques $\tau_d$ (cyan curves) and field-like torques $\tau_f$ (purple curves) as a function of in-plane $\hat{\bold{M}}_\text{Py}$ direction ($\phi$). When the system possesses the mirror plane symmetry, i.e., $\Delta = 0$, the total spin torque must vanish when $\hat{\bold{M}}_\text{Py}~||~\hat{\bold{z}} \times \bold{E}$ (here at $\phi = 90\degree$). Since the magnetization is swept in-plane, the damping-like and field-like torques are the in-plane and out-of-plane torque components respectively, and are fitted to the following form:
\begin{align}
\label{FitEq}
\tau^\text{fit}_{d/f} &= A_{d/f}\cos(\phi + b_{d/f}).
\end{align}
To define the conventional (Y) and unconventional (X) torques, we use these relations:
\begin{align}
\label{DefConv} \tau^Y_{d/f} &= A_{d/f}\cos(b_{d/f}) \\
\label{DefUnconv} \tau^X_{d/f} &= A_{d/f}\sin(b_{d/f}).
\end{align}
Fig. \ref{Fig_Boltzmann}(d) shows the conventional and unconventional torques versus symmetry-breaking parameter $\Delta$. Both the conventional and unconventional field-like torques are greater than the damping-like torques, showing reasonable agreement with the experimental results. While the field-like torque remains dominant under small changes in the parameters of the model, further research is needed to ascertain whether this property is generic for the torque generated by the indirect mechanism.

%
%
%
%

\section{Conclusion}
In summary, we have shown that an unconventional in-plane field-like torque is generated in both ordered (ferrimagnetic) and disordered (paramagnetic) CrPt$_3$. Our spin-torque ferromagnetic resonance and second harmonic Hall measurements reveal that we only observe an unconventional spin-torque component when current is applied along the $[1\overline{1}1]$ and $[\overline{1}11]$ crystal directions. When current is applied along the $[001]$ and $[\overline{1}10]$ crystal directions only conventional torques are observed. Our calculations reveal that symmetry breaking at the CrPt$_3$/Cu interface combined with an indirect non-local spin-orbit torque leads to unconventional spin-torques.

\section{Acknowledgement}
This work was supported as part of Quantum Materials for Energy Efficient Neuromorphic Computing (Q-MEEN-C), an Energy Frontier Research Center funded by the US Department of Energy, Office of Science, Basic Energy Sciences under Award No. DE-SC0019273. KDB acknowledges support from the National Science Foundation under Grants No. DMR-1916275 and DMR-2324203. VA acknowledges support from the National Science Foundation under grants DMR-2105219 and ECCS-2236159.

\section{Data Availability}
The data that support the findings of this study are available from the corresponding author upon reasonable request.

\bibliography{CrPt3Literature}

\end{document}